\documentclass[twocolumn,showpacs,amsmath,amssymb,prb]{revtex4}

\usepackage{graphicx}
\usepackage{dcolumn}
\usepackage{bm}

\begin{document}


\title{Phonon spectrum and soft-mode behavior of MgCNi$_3$}
\author{R. Heid$^{1}$, B. Renker$^{1}$, H. Schober$^{2}$, P. Adelmann$^{1}$,
         D. Ernst$^{1}$, and K.-P. Bohnen,$^{1}$ }
\affiliation{$^{1}$Forschungszentrum Karlsruhe, Institut f\"ur Festk\"orperphysik,
        P.O. Box 3640, D-76021 Karlsruhe, Germany}
\affiliation{$^{2}$Institut Laue-Langevin, BP 156 X, F-38042
       Grenoble Cedex, France }

\date{\today}

\begin{abstract}
Temperature dependent inelastic neutron-scattering measurements of the
generalized phonon density-of-states for superconducting MgCNi$_3$,
T$_c=8$\,K, give evidence for a soft-mode behavior of low-frequency Ni
phonon modes. Results are compared with ab initio density functional
calculations which suggest an incipient lattice instability of the
stoichiometric compound with respect to Ni vibrations orthogonal to the
Ni-C bond direction.

\end{abstract}
\pacs{74.25.Kc,63.20.Kr,78.70.Nx,74.70.Dd}

\maketitle

The recent discovery of superconductivity near 8 K in MgCNi$_3$
\cite{he01} was a surprise since due to the high Ni content a
ferromagnetic ground state would have been expected rather than a
superconducting one. Furthermore it is unusual that for this
perovskite compound the heavy Ni atoms occupy the corners in
Ni$_6$C octahedra. There are speculations that magnetic
interactions might promote superconductivity \cite{rosne02}, however,
measurements of the specific heat come up with the conclusion
that MgCNi$_3$ is a medium to strong coupling BCS superconductor
\cite{lin03,lin03b}.
This compound bears similarity to superconducting nickel
borocarbides where for the Y and Lu compounds an acoustic phonon
branch exhibits a pronounced softening at lower temperatures
\cite{dervenagas,gompf,zares99}, a feature which is supposed to be
connected to superconductivity in these compounds.
Hints for unusual lattice dynamical properties of MgCNi$_3$ have been
inferred from a recent x-ray absorption study, which indicated deviations
of the local atomic structure from the ideal perovskite lattice at
temperatures below 70\,K \cite{ignat03}, and from a very recent linear
muffin-tin orbitals (LMTO) calculation of the harmonic phonons of
MgCNi$_3$ \cite{ignat03b}, which found a dynamical instability of the
stoichiometric compound.
In this paper we will present inelastic neutron scattering (INS)
measurements of the generalized phonon density-of-states (GDOS) as
a function of temperature which show a softening of low-frequency Ni
modes. We will compare our experimental results to ab initio
density functional calculations.

\begin{figure}[t]
\includegraphics[width=0.9\linewidth]{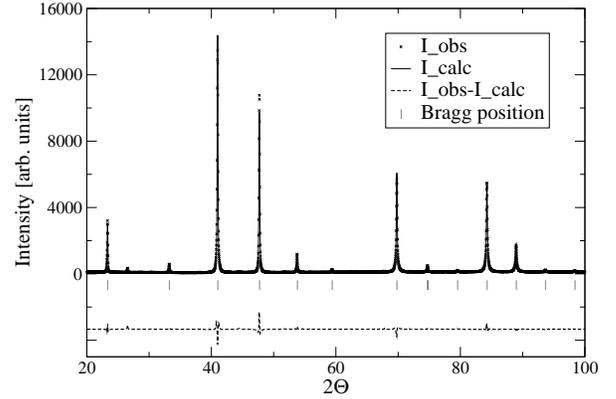}
\caption{Observed (crosses) and calculated (solid line) x-ray
diffraction pattern for the investigated sample. A small impurity
peak at 26.5$^{\circ}$ is due to unreacted graphite. The
refinement (R = 5$\%$) yields a composition of MgC$_{.96}$Ni$_3$
for the cubic perovskite structure, space group Pm3m, with lattice
constant $a$=3.8085(1)\AA\ and with positions of the atoms at: Mg 1a (0,0,0), C 1b
(0.5,0.5,0.5), and Ni 3c (0,0.5,0.5).}
\label{fig1}
\end{figure}

For the synthesis of MgCNi$_3$ samples mixed powders of the
components were pressed into pellets, wrapped into a tantal foil,
sealed in a quartz tube and heated for 2\,h at 900$^{\circ}$C. A
difficulty arises from losses of volatile Mg and C at the
synthesis temperature. With respect to subsequent x-ray analysis
we obtained the best samples for a starting stoichiometry of
Mg$_{1.14}$C$_{1.4}$Ni$_3$ where an impurity peak due to unreacted
graphite was smallest. The T$_c$ of our samples was 
7.5\,K. It could be slightly increased for the starting
concentration C$_{1.5}$ however at the expense of a somewhat
larger contribution of unreacted graphite. Since the exact
stoichiometry might be important for a soft-mode behavior which
is found in the present investigation we show in Fig.~\ref{fig1} a x-ray
diffraction pattern of our sample. 

Our INS measurements were performed at
the HFR in Grenoble, France, on the IN6 and  IN4 time-of-flight
spectrometers. On IN6 a high chopper speed of 12060\,rpm and focusing in
the inelastic region were used to improve the resolution. Due to
the low incident energy of 4.75\,meV at this spectrometer which is
connected to the cold source we profit from a very good
resolution for low-energy excitations. The necessity to work in
energy gain of the scattered neutrons sets limits on the
possibility to work at lower temperatures. For example, for a sample
temperature of 50\,K we could follow excitations up to $\sim$\,30\,meV. The
spectrometer is dedicated to inelastic measurements and a
standard result is the GDOS which is
calculated from the recorded intensities over a scattering region
from 14$^{\circ}$ to 114$^{\circ}$. For the data evaluation we
have applied multi-phonon corrections in a self-consistent
procedure \cite{Schober}. The GDOS implies a weighting of vibrational
modes by $\sigma/m$ (scattering cross-section over the mass; we
used values of 0.46, 0.179, and 0.31 barn/amu  for C, Mg, and Ni,
respectively). From a
plot of the elastic intensities over momentum transfer we do not
obtain any evidence for a structural phase transition within the
investigated temperature region in agreement with previous 
results from structural investigations \cite{he01,huang01}. However,
due to the low incident energy this analysis is limited to
$Q\leq$\,2.2\,\AA$^{-1}$.
To study changes in the dynamics in a larger temperature region,
we have performed supplementary measurements with an incident
neutron energy of 35\,meV in the down-scattering mode over a wide range
of temperatures on the instrument IN4.
These results will be presented in the form of the dynamical susceptibility
$\chi''(\omega)$ which is much closer related to the scattering function
$S(Q,\omega)$ than the GDOS \cite{Schober}.

\begin{figure}[t]
\includegraphics[width=1.0\linewidth]{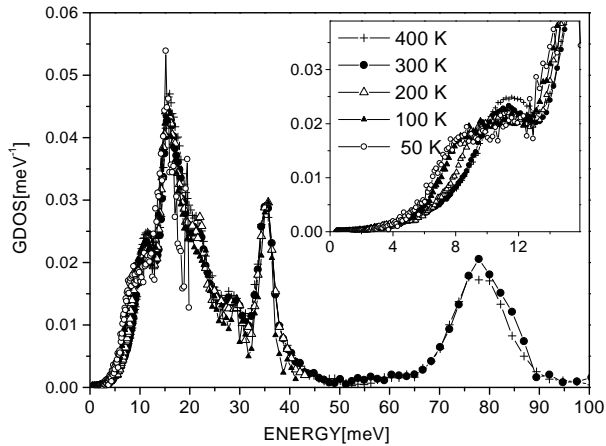}
\caption{The experimental generalized phonon density-of-states
for MgCNi$_3$. A significant soft-mode behavior is found for low-frequency
Ni modes (insert).
} 
\label{fig_gdos}
\end{figure}

Due to the large differences in mass the phonon spectrum of
MgCNi$_3$ (Fig.~\ref{fig_gdos}) decomposes into two well separated parts, a
low-frequency region with predominantly Ni and Mg modes and
vibrations of the light C atom around 80\,meV. From the known
$\sigma/m$ values and an analysis of the areas in the GDOS shown in
Fig.~\ref{fig_gdos} we find for the latter peak an almost perfect
agreement between the theoretical and the experimental spectral
weight implying contributions from three closely neighbored
optical phonon branches. The weight of the sharp and well
separated peak at 35\,meV corresponds well to the weight expected
for three optical Mg modes. It is increased by some 20$\%$ which
suggests a small hybridization with the low-frequency Ni
modes. The region below 33\,meV contains exclusively Ni
contributions. It is well structured and exhibits a very low
frequency maximum at 12\,meV and a strong main peak around
16\,meV.

The 12 zone center optic phonons decompose according to 
3F$_{1u}$\,\,+\,F$_{2u}$ with no Raman active modes. 
Guided by investigations of the related system of quaternary borocarbides
where a soft-mode behavior for superconducting RENi$_2$B$_2$C (RE\,=\,Y,
Lu) has been found we have measured the GDOS of MgCNi$_3$ for a
series of temperatures between 400\,K and 50\,K. 
No changes are found for the peaks at 80 meV and 35 meV which were
attributed to C and Mg vibrations (only a small sharpening of the 35\,meV
peak is registered). 
A pronounced softening, however, occurs for the very low-frequency part
of the Ni vibrations in a region which is dominated by acoustic phonon
branches. The insert in Fig.~\ref{fig_gdos} shows a
magnification. 
The strongest effect is observed between 200\,K and 100\,K.
Only less pronounced changes occur between 100\,K and 50\,K which
is the lowest temperature in our energy-gain measurements.

\begin{figure}[t]
\includegraphics[width=0.7\linewidth]{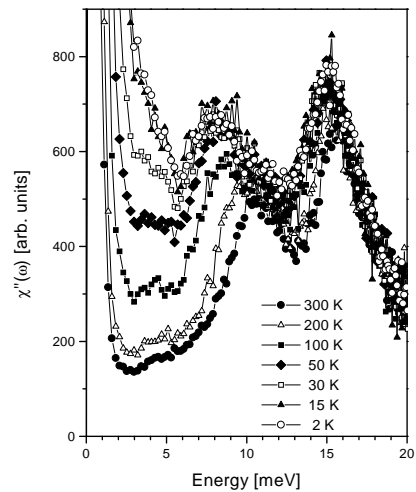}
\caption{The experimental dynamical susceptibility
for MgCNi$_3$ (see text).} 
\label{fig_chi}
\end{figure}

Changes in the dynamics show up clearer in the temperature evolution of
$\chi''(\omega)$ in Fig.~\ref{fig_chi} where a sum of intensities recorded on IN4 for
different scattering angles is shown.
The incident neutron energy was chosen such that the low-energy region below 20\,meV
could still be studied with good resolution.
The strong increase in intensity below 3\,meV is due to elastic scattering contaminations.
Whereas the frequency of the prominent Ni peak remains almost fixed at 16\,meV we can observe
the evolution of the soft-mode intensity.
In agreement with the GDOS the strongest changes are observed around 100\,K and only
minor changes are recorded between 300\,K and 200\,K and between 30\,K
and 2\,K. Additionally
to the GDOS in Fig.~\ref{fig_gdos} it can be seen that some intensity is shifted
into the low-frequency region below 5\,meV.
While the increase in intensity below 10\,meV is in general agreement
with the mode softening a more detailed analysis of its location within
the Brillouin zone (BZ) is not possible for this polycrystalline sample.

\begin{figure}[t]
\includegraphics[width=1.0\linewidth]{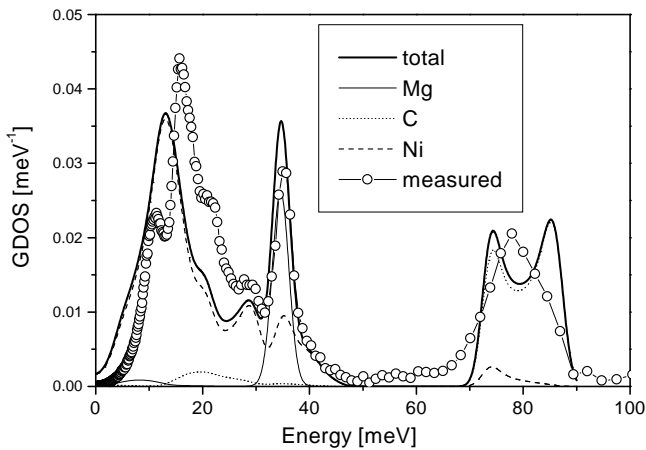}
\caption{Calculated GDOS spectra for stoichiometric MgCNi$_3$.
The proper $\sigma/m$ values have been applied for a comparison
to the experimental data (300\,K spectrum). Contributions from an unstable phonon
branch (see text) have been ignored. 
} 
\label{fig_thgdos}
\end{figure}

For an analysis of our data we have performed first principals density
functional calculations using a mixed basis pseudopotential method.
For Mg a well tested pseudopotential of Troullier-Martins type has been
used \cite{troul91,Pelg}, whereas for C and Ni Bachelet-Hamann-Schl\"uter
type pseudopotentials were constructed \cite{Bachelet}.
The fairly deep $d$ potential of Ni and $p$ potential of C can
be efficiently dealt with by the mixed-basis formalism, where the
valence states are constructed from a combination of localized $s$
and $p$ functions at C sites and localized $d$ functions at Ni sites,
supplemented by plane waves up to a kinetic energy of 24\,Ry.
The local density approximation with the Hedin-Lundqvist parametrization
of the exchange-correlation functional \cite{hedin71} has been used.
Our calculated electronic band structure is in good agreement with
previous density functional studies
\cite{haywa01,szaje01,dugda01,singh01,shim01,shein01,rosne02,kim02}.
The most prominent feature of the electronic density-of-states is the existence
of a strong and sharp van Hove peak less than 100 meV below E$_F$.
It arises from antibonding bands related to the Ni $d$ orbitals, which are
very narrow due the linear Ni coordination by C \cite{singh01}.

\begin{figure}
\includegraphics[width=0.9\linewidth,clip]{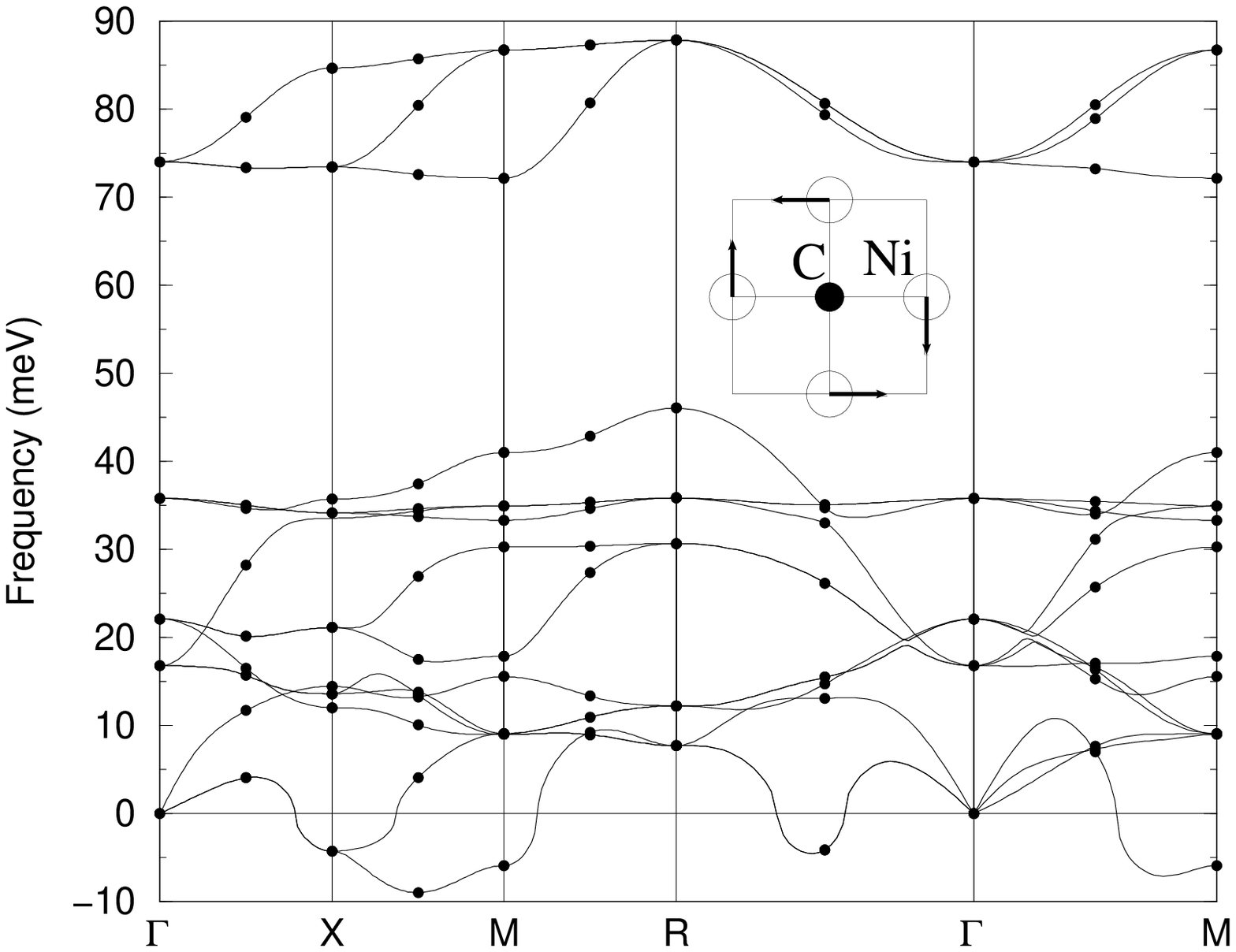}
\caption{Calculated phonon dispersion curves for stoichiometric
MgCNi$_3$ for selected symmetry directions. Black circles denote the
results from linear-response
calculations. Lines are obtained by Fourier interpolation. Negative
frequencies indicate unstable modes.
At the M point, the unstable phonon corresponds to shear elongations of
Ni atoms in the Ni-C planes as indicated in the insert.} 
\label{fig_disp}
\end{figure}

Phonon dispersion curves for stoichiometric MgCNi$_3$ were calculated
within the mixed-basis perturbation approach \cite{heid99b}.
Complete spectra are obtained from a Fourier interpolation of dynamical
matrices calculated on a cubic (4$\times$4$\times$4) q-point mesh.
Results for the optimized lattice constant $a$=3.76\AA\ 
are depicted in Figs.~\ref{fig_thgdos}
and \ref{fig_disp}.
The theoretical GDOS (Fig.~\ref{fig_thgdos}) reproduces many features
seen in the experimental spectra, including the
small maximum near 30\,meV and the shoulder of the acoustic spectrum
at 20\,meV.
The acoustic modes are predominantly of Ni character with only a small
hybridization of C modes.
The sharp peak near 35\,meV arises from three almost dispersionless Mg
branches with a small admixture of Ni vibrations.

The most prominent feature of the calculated dispersion curves
(Fig.~\ref{fig_disp}) is, however, the occurrence of unstable modes
suggesting that the stoichiometric compound is dynamically unstable in
the harmonic approximation.
This instability appears over a large area in reciprocal space and is
most pronounced near the zone boundary points X and M (with coordinates
(100) and (110) in units of $\pi$/a, respectively).
Inspection of the displacement vectors reveals that all unstable modes involve 
Ni vibrations orthogonal to the Ni-C bond directions.
This behavior is most evident at the M point where the unstable mode
represents a pure vibration of the Ni atoms with all other atoms
at rest.
Its elongation pattern, shown in the insert of Fig.~\ref{fig_disp},
induces shear distortions in the planar Ni-C sublattice.
This extended phonon instability may be the origin of the apparent
downshift of the acoustic part below 20\,meV in the calculated spectrum
by 3\,meV with respect to the experimental one (see Fig.~\ref{fig_thgdos}).

The presence of the van Hove peak close to E$_F$ requires care in
performing the BZ summations in the calculations.
The presented results were obtained with cubic (16$\times$16$\times$16)
Monkhorst Pack k-point meshes and a Gaussian broadening of 0.1\,eV.
Convergence studies showed that this is sufficient for most of the phonon
modes to be converged within 0.5\,meV, but the low-frequency Ni modes
are more sensitive to the BZ sampling.
Frozen-phonon studies of the unstable M-point mode using a variety of
k meshes and Gaussian smearings have confirmed that this instability is
not the result of an insufficient BZ sampling.
Furthermore, a similar frozen-phonon calculation performed with the
projector-augmented plane wave approach as implemented in the Vienna ab-initio
simulation package (VASP-PAW)
\cite{vasp,vasp2,vasp3,vasppaw} reproduced the instability, excluding the possibility  
that it is an artefact of the pseudopotential approximation.

An instability of Ni modes has been also found in a very recent
ab initio investigation of the lattice dynamics and electron-phonon
coupling (EPC) of MgCNi$_3$ using the LMTO method
\cite{ignat03b}.
In contrast to our calculations, however, the instability did not appear at
high-symmetry points at the BZ boundary. 
While both methods agree with respect to the frequency range for the
Ni and C modes, the LMTO calculations predict significantly higher
frequencies for the Mg modes (at $\approx$45\,meV) at variance with our
experimental and theoretical results.

The instability of the harmonic lattice dynamics of stoichiometric
MgCNi$_3$ raises the question of the true low-temperature structure of
this superconductor.
Some insight can be gained from frozen-phonon studies for the unstable
mode.
For the shear mode at M we found a very shallow double-well potential,
with a minimum of less than 1\,meV at a displacement of only 0.025\,\AA.
In the previous LMTO study, double-well potentials of similar magnitude
have been extracted from frozen-phonon calculations corresponding to
lower-symmetry points \cite{ignat03b}.
The shallowness of the double well suggests that the instability is not
strong enough to induce a long-ranged structural distortion, but that
the high-symmetry cubic structure is stabilized dynamically.
The presence of strong anharmonic motions of atoms in the vicinity of a
structural instability could be of relevance for superconductivity, as it may
significantly enhance the EPC strength.
For MgCNi$_3$, a contribution of 37\% coming from the anharmonic modes
to the total EPC constant has been estimated on
the basis of the LMTO results \cite{ignat03b}.
A complete calculation of the EPC strength, however, is a formidable task
as it requires to take into account the full anharmonic lattice dynamics
as well as anharmonic corrections to the EPC.
This is further complicated by the fact that the local distortions
are expected to strongly modify the van Hove singularity and thus the
electronic structure in the vicinity of the Fermi energy \cite{ignat03b}.

The picture of a dynamically stabilized cubic structure is consistent with
published structural investigations where no indications of any phase
transition have been found, and with the temperature dependence
of the elastic scattering intensity observed in our IN6 spectra.
In a recent x-ray absorption fine structure measurement, an unusually broad Ni-Ni
pair-correlation function has been observed below 70\,K \cite{ignat03}.
It was interpreted as the result of local distortions of the Ni octahedra
predominantly perpendicular to the Ni-C bond with displacements of less
than 0.05\,\AA, in agreement with the frozen-phonon results.

In conclusion, studies of the GDOS have provided a clear insight in the
lattice vibrations of superconducting MgCNi$_3$ which can be looked at
as a 3D analogue of the nickel borocarbides.
For both kinds of compounds a remarkable soft-mode behavior in the
low-frequency region ($\sim$\,8\,meV) is observed.
For the superconducting borocarbides where single crystals could be
grown it was shown that the mode softening is restricted to a narrow
region in reciprocal space and is sensitive to the formation of the
superconducting gap \cite{zares99}.
In contrast, the present results suggest that the softening phenomena
in MgCNi$_3$ is not directly linked to the superconducting state, but
has its origin in an incipient dynamical instability of specific Ni
vibrations which occurs widespread in reciprocal space.
It cannot be excluded, however, that lattice defects play an important
role in stabilizing the real structure.
Further studies are required to clarify the question whether the
apparently large stability range of the MgC$_{1-x}$Ni$_3$ phase is
connected to the observed dynamical peculiarities.

\end{document}